\documentclass[conference]{IEEEtran}

\usepackage{amsmath,epsfig,amssymb,verbatim}
\usepackage{cite}
\usepackage[caption=false]{subfig}

\def\BibTeX{{\rm B\kern-.05em{\sc i\kern-.025em b}\kern-.08em
    T\kern-.1667em\lower.7ex\hbox{E}\kern-.125emX}}

\begin{document}

\title{Coverage Analysis for 3D Terahertz Communication Systems with Blockage and Directional Antennas}

\author{\IEEEauthorblockN{Akram Shafie, Nan Yang, Zhuo Sun, and Salman Durrani}
\IEEEauthorblockA{Research School of Electrical, Energy and Materials Engineering\\Australian National University, Canberra, ACT 2600, Australia}
\IEEEauthorblockA{Email: akram.shafie@anu.edu.au, nan.yang@anu.edu.au, zhuo.sun@anu.edu.au, salman.durrani@anu.edu.au}}

\markboth{Submitted to IEEE ICC 2020 Workshop}{Akram \MakeLowercase{\textit{et
al.}}: Coverage Analysis for 3D Terahertz Communication Systems With Blockage and Directional Antennas}

\maketitle

\begin{abstract}

The scarcity of spectrum resources in current wireless communication systems has sparked enormous research interest in the terahertz (THz) frequency band. This band is characterized by fundamentally different propagation properties resulting in different interference structures from what we have observed so far at lower frequencies. In this paper, we derive a new expression for the coverage probability of downlink transmission in THz communication systems within a three-dimensional (3D) environment. First, we establish a 3D propagation model which considers the molecular absorption loss, 3D directional antennas at both access points (APs) and user equipments (UEs), interference from nearby APs, and dynamic blockages caused by moving humans. Then, we develop a novel easy-to-use analytical framework based on the dominant interferer analysis to evaluate the coverage probability, the novelty of which lies in the incorporation of the instantaneous interference and the vertical height of THz devices. Our numerical results demonstrate the accuracy of our analysis and reveal that the coverage probability significantly decreases when the transmission distance increases. We also show the increasing blocker density and increasing AP density impose different impacts on the coverage performance when the UE-AP link of interest is in line-of-sight. We further show that the coverage performance improvement brought by increasing the antenna directivity at APs is higher than that brought by increasing the antenna directivity at UEs.
\end{abstract}

\begin{IEEEkeywords}
Terahertz communication, coverage, 3D modeling, directional antennas, dynamic blockage.
\end{IEEEkeywords}

\section{Introduction}\label{sec:introduction}

Terahertz (THz) communication has been envisioned as a highly promising paradigm to support hyper-fast data transmission with ultra-high data rate in the sixth-generation (6G) wireless networks \cite{2018MagLastMeterIndoor}. The rationale behind exploring THz communication is to alleviate the spectrum scarcity and break the capacity limitation of contemporary wireless networks. In particular, the ultra-wide THz band ranging from 0.1 to 10 THz provides a huge potential to realize 6G applications which demand multi-terabits per second (Tb/s) data transmission, such as ultra-fast wireless local area networks and wireless virtual/augmented reality. Notably, such demand is beyond the capability of emerging millimeter wave (mmWave) communication which is anticipated to be used in the near future \cite{2018MagCombatDist}.

Despite its high promise, the THz band encounters numerous new and pressing challenges that have never been seen at lower frequencies. For example, the THz band suffers from very high spreading loss and molecular absorption loss which profoundly decreases the THz transmission distance \cite{A0}. Moreover, high reflection and scattering losses attenuate the non-line-of-sight (NLOS) rays significantly, triggering the need for line-of-sight (LOS) link for reliable transmission. Furthermore, THz signal propagation is highly vulnerable to blockages that are caused by moving humans and indoor constructions (e.g., walls and furnitures) \cite{2017M5}. All these challenges lead to unique propagation environment at the THz band, which motivates the design and development of new communication paradigms and novel signal processing tools.

Multiple antennas are possible to be integrated into THz transmitters such that super-narrow directional beams are formed to overcome severe path loss \cite{NanHangGlobeComBeamforming}. The use of such highly directional antennas may lead to the noise-limited regime of THz communication. However, the increase in network densification, the use of advanced networking mechanisms such as pico/femto cells, and direct device-to-device communication are likely to increase the impact of interference on THz communication systems \cite{2017M5}. Therefore, the evaluation of the reliability of THz communication systems in the presence of interference is an important research problem.

Coverage probability is a widely used performance metric to quantify reliability. Conventionally, in sub-6 GHz and mmWave communication systems, the coverage probability in the presence of interference has been derived with the aid of Laplace transform-based analysis  \cite{StochGeozRef1, LaplaceIntRef2}. However, it is fundamentally difficult to apply this approach in the THz band because of two reasons. First, there is a lack of closed-form expression for the Laplace transform of the interference from a single THz node, due to the exponential term in the THz channel. Second, the distance dependant blockage effect leads to non-uniform interferers  \cite{2017M5}. As a result, the studies on the coverage probability in the presence of interference in the THz band are limited, except for \cite{2017M12,2017N3,2017M5}. Constrained by the aforementioned reasons, \cite{2017M5} derived the first few moments of interference and signal-to-interference-plus-noise ratio (SINR). In \cite{2017N3}, the evaluation of the coverage probability used the average interference instead of the instantaneous interference. Although the instantaneous interference was considered in \cite{2017M12}, it made an assumption that the channel is interference-limited.
In addition, it is worthwhile to note that \cite{2017M12,2017N3,2017M5} focused on a two-dimensional (2D) environment only, which implies that the vertical height of THz devices was not examined. However, this vertical height introduces considerable complexity into the analysis, and may greatly impact the reliability performance of THz communication systems.

In this paper, we develop a novel easy-to-use analytical framework using the tools of stochastic geometry to evaluate the coverage probability of THz communication systems in a three-dimensional (3D) environment. For the system, we establish a 3D propagation model where we consider the molecular absorption loss which is unique in the THz band, 3D directional antennas at both the transmitters and the receivers, the interference from nearby transmitters, and dynamic blockage caused by moving humans. Under such consideration, we derive new expressions for the coverage probability of downlink transmission using the dominant interferer analysis. Here, the coverage probability is defined as the probability that the SINR at the target receiver is larger than a predefined threshold.
Different from the current literature, the proposed analytical framework incorporates the instantaneous interference as well as the vertical height of THz transmitters and receivers. Aided by numerical results, we demonstrate that our analysis is accurate. We also find that the coverage probability significantly deteriorates when the transmission distance becomes large. Moreover, we find that an increase in the density of blockers leads to a reduction in coverage performance, but slightly improves the coverage when the transmission link of interest is in LOS. Furthermore, we find that the denser deployment of transmitters significantly reduces the coverage performance, while this reduction can be compensated for by improving the antenna directivity at transmitters and receivers. Additionally, we find that the coverage performance gain brought by the increase in antenna directivity at transmitters is higher than that brought by the increase in antenna directivity at receivers.

\section{System Model}\label{sec:system_model}

\begin{figure}[!t]
\centering
\includegraphics[width=0.95\columnwidth]{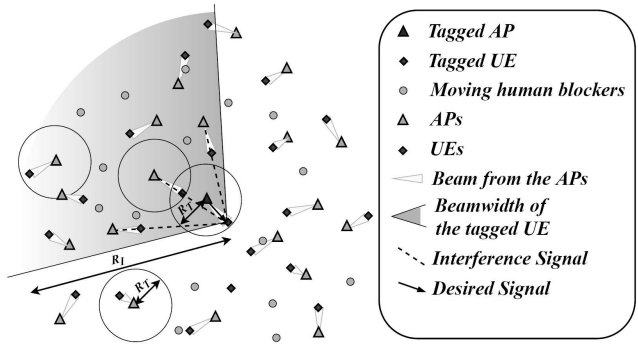}
\caption{Illustration of the top view of the 3D THz communication system.}\label{Fig:SystemModel}
\end{figure}

Fig.~\ref{Fig:SystemModel} depicts the top view of the 3D THz communication system considered in this work. We assume that the THz APs are of fixed height $h_{\textrm{A}}$ and their locations follows a Poisson point process (PPP) in $\mathbb{R}^{2}$ with the density of $\lambda_{\textrm{A}}$. We also assume that user equipments (UEs), all of which are of fixed height $h_{\textrm{U}}$, are distributed uniformly within the circle with the radius $R_{\textrm{T}}$ centered at each AP. Although multiple UEs may exist in each circle, we assume that each AP in the system associates with one UE only. Among the UE-AP pairs, we randomly select an arbitrary pair and denote the UE and the AP in this pair as the tagged UE and tagged AP, respectively. This allows us to characterize the downlink performance at the tagged UE. We assume that all the UE-AP pairs share the same frequency channel; hence, apart from the tagged AP, all the other APs in the system act as ``interferers'' to the tagged UE.

Humans moving in the area of the considered system can act as blockers. Specifically, they can potentially block the desired signals from the tagged AP to the tagged UE, as well as the interference signals from other APs to the tagged UE. We model these humans as cylinders with the radius $r_{\textrm{B}}$ and the height $h_{\textrm{B}}$ \cite{2019NN1}, and their location follows another PPP with the density of $\lambda_{\textrm{B}}$. Furthermore, we assume that the mobility of humans follows the random directional model (RDM). According to this model, if a blocker is moving in the area $\mathbb{R}^2$, the probability density function (PDF) of its location is uniform over time \cite{RDM1}. As such, at any given time instant, the location of blockers forms a PPP with the same density of $\lambda_{\textrm{B}}$. Considering the practical aspects, we assume that $h_{\textrm{A}} > h_{\textrm{B}} > h_{\textrm{U}}$.

\subsection{Propagation Model}

The signal propagation at THz frequencies is determined by spreading loss and molecular absorption loss \cite{A0}. As such, the received power of an arrival ray in the 3D THz channel is given by
\begin{align}\label{Equ:Pr}
P_{r}(x)
=&\varrho\; d(x)^{-2}e^{-K(f)d(x)},
\end{align}
where $\varrho\triangleq P_{\textrm{T}}G_{\textrm{A}}G_{\textrm{U}}c^{2}/\left(4\pi f\right)^{2}$, $P_{\textrm{T}}$ is the transmit power, $G_{\textrm{A}}$ and $G_{\textrm{U}}$ are the antenna gains at the AP and the UE, respectively, $c$ is the speed of light, $f$ is the operating frequency, $x$ and $d(x)$ are the 2D and 3D propagation distances between the UE and the AP, respectively, with $d(x)=\sqrt{(h_{\textrm{A}} - h_{\textrm{U}})^2 +x^2}$, and $K(f)$ is the frequency-dependent molecular absorption loss coefficient of the transmission medium.

\begin{figure}[!t]
\centering
\includegraphics[height=1.5in]{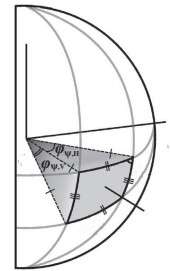}
\caption{3D antenna radiation pattern.}\label{Fig:3Dbeam}
\end{figure}

In this work, we assume that 3D beams are utilized at the APs and the UEs.
This is a reasonable assumption since directional antennas are expected to be used at both the transmitter and the receiver in THz communication systems to compensate for the severe path loss \cite{2018MagCombatDist}. We model the 3D beam of the THz devices with a pyramidal zone for its given horizontal beamwidth, $\varphi_{\Psi,\textrm{H}}$, and the vertical beamwidth, $\varphi_{\Psi,\textrm{V}}$, as shown in Fig.~\ref{Fig:3Dbeam}, where $\Psi \in \{\textrm{A},\textrm{U}\}$. If $G_{\Psi}$ is the antenna gain corresponding to $\varphi_{\Psi,\textrm{H}}$ and $\varphi_{\Psi,\textrm{V}}$, then from \cite{2019NN1} we can express $G_{\Psi}$ as
\begin{equation}\label{Equ:G}
G_{\Psi}=\pi
\left(\arcsin\left(\tan\left(\frac{\varphi_{\Psi,\textrm{H}}}{2}\right) \tan\left(\frac{\varphi_{\Psi,\textrm{V}}}{2}\right)\right)\right)^{-1}.
\end{equation}
In addition, in this work we focus on the LOS rays of THz signals. When signals are propagated in the THz band, the direct ray dominates the received signal energy, due to the high directional nature and the high reflection loss of THz beams \cite{A0}.

\subsection{Blockage}

\begin{figure}[!t]
\centering\subfloat[Side view\label{1a}]{\includegraphics[clip,height=1.4in]{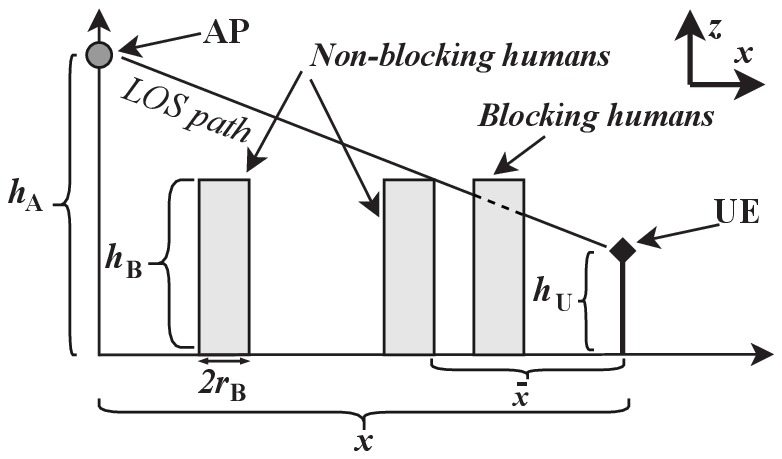}}\hfill
\subfloat[Top view\label{1b}]{\includegraphics[clip,height=1.4in]{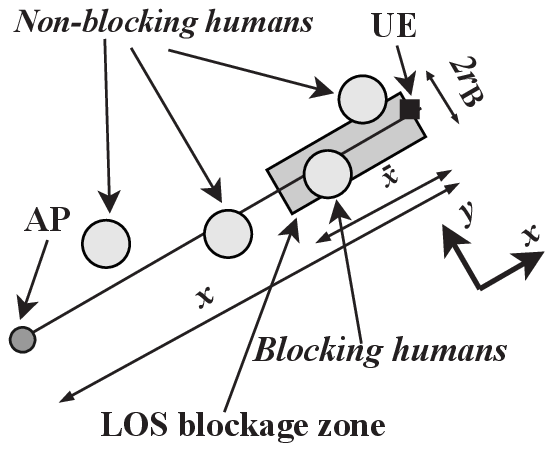}}
\caption{Illustration of a single UE-AP link in the presence of blockers.}\label{Fig:BlockModel}
\end{figure}

The LOS link between an AP and the UE is blocked if at least one blocker appears in the LOS blockage zone of the UE-AP link. For an UE-AP link with a 2D distance of $x$, this area can be approximated by a rectangle between the UE and the AP with sides of $2r_{\textrm{B}}$ and $\bar{x}$, as shown in Fig. \ref{Fig:BlockModel}, where
\begin{equation}\label{Equ:dx}
\bar{x}=\frac{h_{\textrm{B}}-h_{\textrm{U}}}{h_{\textrm{A}}-h_{\textrm{U}}}x+r_{\textrm{B}}.
\end{equation}
Therefore, the LOS probability of the link is same as the void probability of the Poisson process in the LOS blockage zone, which is given by
\begin{equation}\label{Equ:pL}
p_{\textrm{L}}(x)=e^{-2\lambda_{\textrm{B}}r_{\textrm{B}}\bar{x}}=\zeta e^{-\eta x},
\end{equation}
where $\zeta = e^{-2\lambda_{\textrm{B}}r_{\textrm{B}}^2}$ and $\eta = 2\lambda_{\textrm{B}}r_{\textrm{B}}(h_{\textrm{B}}-h_{\textrm{U}})/(h_{\textrm{A}}-h_{\textrm{U}})$.
We clarify that the analysis herein is performed aimed at an open office environment; therefore, only one type of blockers, i.e., human blockages, are considered.
We note that the blockages caused by indoor constructions (e.g., walls and furnitures) may also need to be considered when characterizing a more generalized indoor THz communication environment.
\subsection{Calculation of $R_{\textrm{T}}$}

Recall that UEs are distributed uniformly within the circle with radius $R_{\textrm{T}}$ centered at each AP and each AP associates with one UE only. Here, it needs to be ensured that the signal-to-noise ratios (SNRs) of all the associated UEs are above their predefined threshold, denoted by $\tau$, when the signal at each associated UE from its corresponding AP is not blocked. To this end, the value of $R_{\textrm{T}}$ is determined as a function of the propagation model, transmit power, and antenna gains, which is given by
\begin{equation}\label{Equ:RT}
R_{\textrm{T}}=\sqrt{\left(\frac{2}{K(f)} W\left[\frac{K(f)}{2}\sqrt{\frac{\varrho}{\sigma^{2}\tau}}\right]\right)^2-(h_{\textrm{A}}-h_{\textrm{U}})^2},
\end{equation}
where $\sigma^2$ is the additive white Gaussian noise (AWGN) power in the transmission window of interest and $W\left[\cdot\right]$ is the Lambert \textit{W}-function. The derivation of $R_{\textrm{T}}$ is given in Appendix \ref{app:Derive_RT}.

\section{Coverage Probability Analysis}\label{sec:analysis}

In this section, we derive the coverage probability of downlink transmission at the tagged UE using dominant interferer analysis while considering both blockage and directional antennas.

Let us denote $x_{i}$ as the distance from an AP, i.e., AP$_{i}$, to the tagged UE, where $i=0,1,2,\ldots$. Specifically, AP$_{0}$ is referred to as the tagged AP.
By considering LOS blockage, the coverage probability at the tagged UE, $p_{c}(x_{0})$, is expressed as
\begin{equation}\label{Equ:pc1}
p_{c}(x_{0}) = p_{\textrm{L}}(x_{0}) p_{c,\textrm{L}}(x_{0}),
\end{equation}
where $p_{\textrm{L}}(x_{0})$ is the LOS probability calculated in
\eqref{Equ:pL} and $p_{c,\textrm{L}}(x_{0})$ is the probability of the SINR at the tagged UE being larger than $\tau$, when the link between the tagged UE and the tagged AP is in LOS. In particular, $p_{c,\textrm{L}}(x_{0})$ is written as
\begin{align}\label{Equ:pcorg}
p_{c,\textrm{L}}(x_{0}) &=\mathbb{P}\;\left[\textrm{SINR}\vert_{\textrm{LOS}} \geq \tau\right] \notag \\
&=\mathbb{P}\left[\frac{P_{r}(x_{0}) }{\sigma^{2}+\sum I} \geq \tau\right],
\end{align}
where $\sum I$ denotes the aggregated interference at the tagged UE. From \eqref{Equ:pcorg}, it is evident that the analysis of $\sum I$ is essential to derive $p_{c,\textrm{L}}(x_{0})$. To this end, the APs which contribute to $\sum I$ at any given time instant need to be identified. Hence, we denote $\Phi$ as the set of APs which contribute to the interference at the tagged UE and will characterize $\Phi$ in the next subsection.

\subsection{Characterization of $\Phi$} \label{SubSec:phi}

\begin{figure}[!t]
\centering\subfloat[Side view\label{2a}]{\includegraphics[clip,width=1\columnwidth]{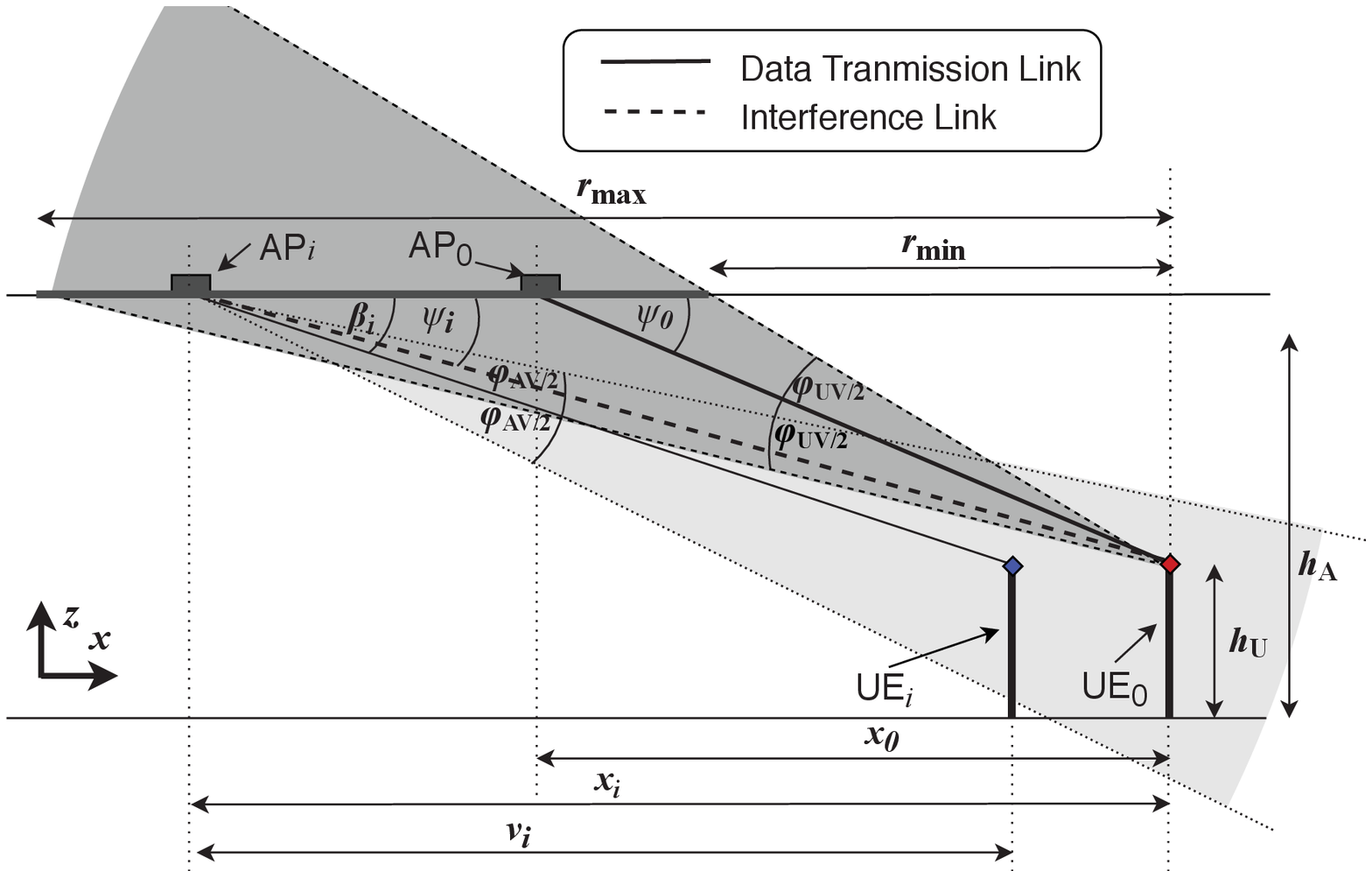}}\hfill
\subfloat[Top view\label{2b}]{\includegraphics[clip,height=2in]{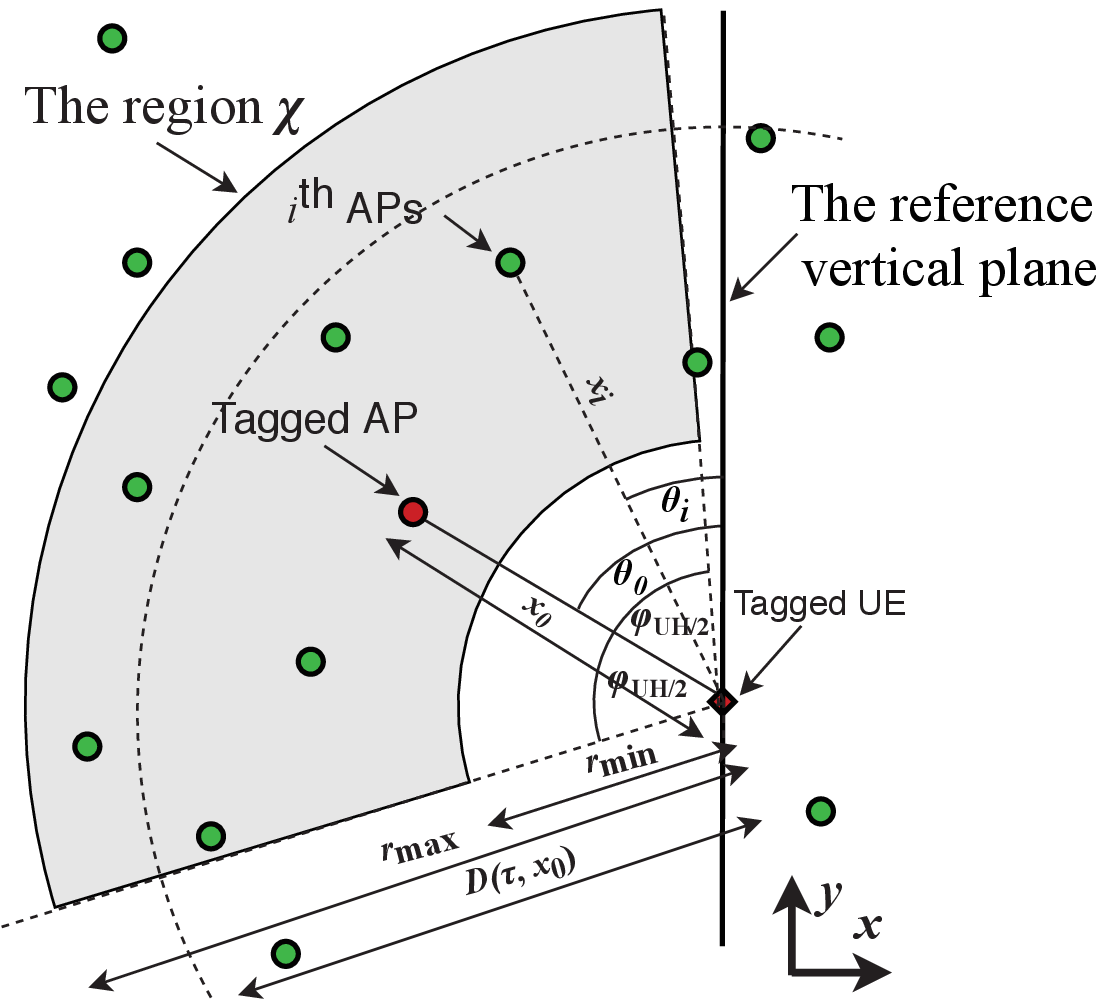}}
\caption{Illustration of a single UE-AP link in the presence of interferers.}\label{Fig:InfModel}
\end{figure}

By examining the characteristics of the considered THz communication system, we point out that several conditions need to be satisfied for an AP to contribute to the aggregated interference at the tagged UE. These conditions are:
\begin{enumerate}
\item The AP is within the horizontal beamwidth of the tagged UE;
\item The AP is within the vertical beamwidth of the tagged UE;
\item The tagged UE is within the horizontal beamwidth of the AP;
\item The tagged UE is within the vertical beamwidth of the AP;
\item The link between the AP and the tagged UE is not blocked by moving humans.
\end{enumerate}
In the following, we find the APs apart from the tagged AP which satisfy the aforementioned conditions.

We denote $\theta_{i}$ and $\psi_{i}$ as the angles that the link between AP$_{i}$ and the tagged UE form with a given reference vertical plane and the horizontal plane, respectively,
as shown in the Fig. \ref{Fig:InfModel}. Therefore, for AP$_{i}$ to satisfy \textit{Condition 1)}, $\theta_{i}$ needs to satisfy
\begin{equation}\label{Equ:Cond1}
\theta_{0}-\frac{\varphi_{\textrm{U,H}}}{2}\leq\theta_{i}\leq \theta_{0}+\frac{\varphi_{\textrm{U,H}}}{2}.
\end{equation}
Also, for AP$_{i}$ to satisfy \textit{Condition 2)}, $\psi_{i}$ needs to satisfy
\begin{equation}\label{Equ:Cond2}
\psi_{0}-\frac{\varphi_{\textrm{U,V}}}{2}\leq\psi_{i}\leq\psi_{0}+\frac{\varphi_{\textrm{U,V}}}{2}.
\end{equation}
Based on the knowledge of geometry, \eqref{Equ:Cond2} leads to
\begin{equation}\label{Equ:Cond22}
r_{\textrm{min}}\leq x_{i}\leq r_{\textrm{max}},
\end{equation}
where
\begin{align}\label{Equ:rmax}
r_{\textrm{max}}=\begin{cases}
\frac{(h_{\textrm{A}}{{-}}h_{\textrm{U}})\left( x_{0}+ (h_{\textrm{A}}{-}h_{\textrm{U}}) \tan\left(\frac{\varphi_{\textrm{U,V}}}{2}\right)\right) }{(h_{\textrm{A}}{-}h_{\textrm{U}}){-}x_{0} \tan(\frac{\varphi_{\textrm{U,V}}}{2})}, &\textrm{if~}\psi_{0}\geq \frac{\varphi_{\textrm{U,V}}}{2},\\
\infty, &\textrm{otherwise}
\end{cases}
\end{align}
and
\begin{align}\label{Equ:rmin}
r_{\textrm{min}}=\begin{cases}
\frac{(h_{\textrm{A}}{-}h_{\textrm{U}})\left( x_{0}{-} (h_{\textrm{A}}{-}h_{\textrm{U}}) \tan\left(\frac{\varphi_{\textrm{U,V}}}{2}\right)\right) }{(h_{\textrm{A}}{-}h_{\textrm{U}})+x_{0} \tan(\frac{\varphi_{\textrm{U,V}}}{2})}, &\textrm{if~}\psi_{0}\leq \frac{\pi{-}\varphi_{\textrm{U,V}}}{2},\\
0, &\textrm{otherwise},
\end{cases}
\end{align}
the derivation of which is presented in Appendix \ref{app:Derive_rmax_rmin}. We clarify that $r_{\textrm{max}}$ and $r_{\textrm{min}}$ depend on the distance between the tagged AP and the tagged UE, i.e., $x_{0}$. Accordingly, we denote the region around the tagged UE which satisfies \eqref{Equ:Cond1} and \eqref{Equ:Cond22} by $\chi$ as shown in Fig.~\ref{2b}, where
\begin{equation}\label{Equ:CondX}
\chi=\left\{(x,\theta),x\in\left[r_{\textrm{min}},r_{\textrm{max}}\right],
\theta\in\left[\theta_{0}-\frac{\varphi_{\textrm{U,H}}}{2},\theta_{0}
+\frac{\varphi_{\textrm{U,H}}}{2}\right]\right\}.
\end{equation}

To investigate \textit{Conditions 3)} and \textit{4)}, we denote $p_{\textrm{H,H}}(x_{i})$ and $p_{\textrm{H,V}}(x_{i})$ as the probabilities of the tagged UE being within the horizontal and vertical beamwidths of AP$_i$, respectively. Mathematically, $p_{\textrm{H,H}}(x_{i})$ is given by
\begin{align}\label{Equ:pHH}
p_{\textrm{H,H}}(x_{i})=\frac{\varphi_{\textrm{A,H}}x_{i}}{2\pi x_{i}}=\frac{\varphi_{\textrm{A,H}}}{2\pi}
\end{align}
and $p_{\textrm{H,V}}(x_{i})$ is given by \eqref{Equ:pHV} on the next page,
\begin{figure*}[!t]
\normalsize
\begin{align}\label{Equ:pHV}
p_{\textrm{H,V}}(x_{i})=\begin{cases}
\frac{(h_{\textrm{A}}-h_{\textrm{U}})^{2}}{R_{\textrm{T}}^{2}}\left[\cot^2\left(\psi_{i}-\frac{\varphi_{\textrm{A,V}}}{2}\right)
-\cot^2\left(\psi_{i}+\frac{\varphi_{\textrm{A,V}}}{2}\right)\right], & 0\leq x_{i}\leq x_{\mu},\\
1-\frac{(h_{\textrm{A}}-h_{\textrm{U}})^{2}}{R_{\textrm{T}}^{2}}\cot^{2}\left(\psi_{i}+\frac{\varphi_{\textrm{A,V}}}{2}\right), & x_{\mu}<x_{i}<x_{\nu},\\
0, & x_{i} \geq x_{\nu}.\end{cases}
\end{align}\vspace{-0.5em}
\hrulefill
\end{figure*}
where $x_{\mu}= (h_{\textrm{A}}{-}h_{\textrm{U}})\cot\left(\textrm{min}\left\{\frac{\pi}{2},\bar{\psi}{{+}}\frac{\varphi_{\textrm{A,V}}}{2}\right\} \right)$, and $x_{\nu}=(h_{\textrm{A}}{-}h_{\textrm{U}})\cot \left(\textrm{max}\left\{0,\bar{\psi}-\frac{\varphi_{\textrm{A,V}}}{2}\right\}\right)$ with $\bar{\psi} = \arctan\left(\frac{h_{\textrm{A}}{-}h_{\textrm{U}}}{R_{\textrm{T}}}\right)$ \cite{2019NN1}. The proof of \eqref{Equ:pHV} is given in Appendix \ref{app:Derive_15}.

Finally, the non-blocking probability of the link between AP$_i$ and the tagged UE, i.e.,  the probability for \textit{Condition 5)}, is calculated using \eqref{Equ:pc1}. Therefore, considering \eqref{Equ:CondX},  \eqref{Equ:pHH}, \eqref{Equ:pHV}, and \eqref{Equ:pc1}, we conclude that all the APs in the region $\chi$ contribute to the aggregated interference with the probability of $p_{\epsilon}(x_{i})$. Mathematically, $p_{\epsilon}(x_{i})$ is given by
\begin{equation}\label{Equ:pep}
p_{\epsilon}(x_{i})=p_{\textrm{H,H}}(x_{i})p_{\textrm{H,V}}(x_{i})p_{\textrm{L}}(x_{i}).
\end{equation}
Such APs constitute the set $\Phi$.

\subsection{Dominant Interferer Analysis}\label{subSec:Dom}

In this work, we use the dominant interferer analysis to examine the coverage probability. In doing so, we partition the APs which contribute to the aggregated interference at the tagged UE into two subsets: \textit{dominant} and \textit{non-dominant interferers} \cite{DominantD2D}.
We define an interferer as a \textit{dominant interferer} if it causes outage at the tagged UE when none of the other interferers contribute to the aggregated interference. Moreover, we define an interferer as a \textit{non-dominant interferer} if it cannot cause outage by itself.
Dominant interferer analysis assumes that the presence of any combination of \textit{non-dominant interferers} cannot lead to the outage. This is a reasonable assumption in THz communication systems since the aggregated interference from distant interferers is minimal in such systems, due to the following reasons.
First, the probability of distant interferers causing interference at the tagged UE is very low, due to the use of directional antennas at the UEs and the APs and the fact that the LOS blockage exponentially increases with distance. Second, the interference power from a distant interferer is very small due to the exponential power decay as a result of the molecular absorption loss. We will validate the feasibility of this assumption in Section \ref{sec:numerical}.

By using the dominant interferer analysis, $ p_{c,\textrm{L}}(x_{0}) $ in \eqref{Equ:pcorg} can be interpreted as the probability that no interferers is a \textit{dominant interferer}, when the link between the tagged UE and the tagged AP is LOS. Therefore, $p_{c,\textrm{L}}(x_{0})$ is written as
\begin{align}\label{Equ:pclx0X_Step1}
p_{c,\textrm{L}}(x_{0})
&=\mathbb{P}\left[\frac{P_{r}(x_{0})}{\sigma^{2}+\sum_{\Phi}I}\geq\tau\right]\notag\\
&=\mathbb{P}\left[\sum_{\Phi}\varrho\: d(x_{i})^{-2}e^{-K(f)d(x_{i})}\leq
\frac{P_{r}(x_{0})-\tau\sigma^{2}}{\tau}\right].
\end{align}
We then denote AP$_{i_{c}}$ as the closest interferer of the tagged UE which satisfies the five conditions stated in Section \ref{SubSec:phi}. Following the fact that only the interference from the closest interferer is considered, we obtain
\begin{align}\label{Equ:pclx0X_Step2}
&p_{c,\textrm{L}}(x_{0})
\leq\mathbb{P}\left[\varrho\; d(x_{i_{c}})^{-2}e^{-K(f)d(x_{i_{c}})}\leq
\frac{P_{r}(x_{0})-\tau\sigma^{2}}{\tau}\right]\notag\\
&=\mathbb{P}\left[\frac{K(f)d(x_{i_{c}})}{2}e^{\frac{K(f)d(x_{i_{c}})}{2}}\geq
\frac{K(f)}{2}\sqrt{\frac{\varrho\;\tau}{P_{r}(x_{0})-\tau\sigma^{2}}}\right].
\end{align}
Next, we apply the definition of the Lambert \textit{W}-function to \eqref{Equ:pclx0X_Step2}, which leads to
\begin{align}\label{Equ:pclx0X_Step3}
p_{c,\textrm{L}}(x_{0})
&=\mathbb{P}\left[\frac{K(f)d(x_{i_{c}})}{2}\geq W\left[\frac{K(f)}{2}\sqrt{\frac{\varrho\;\tau}
{P_{r}(x_{0})-\tau\sigma^{2}}}\right]\right]\notag\\
&=\mathbb{P}\left[d(x_{i_{c}})\geq\frac{2}{K(f)}W\left[\frac{K(f)}{2}\sqrt{\frac{\varrho\;\tau}
{P_{r}(x_{0})-\tau\sigma^{2}}}\right]\right]\notag\\
&=\mathbb{P}\left[x_{i_{c}}\geq D(\tau, x_{0})\right],
\end{align}
where $D(\tau, x_{0})$ is the distance from the tagged UE to the boundary of the region around the tagged UE where \textit{dominant interferers} can exist. Mathematically, $D(\tau, x_{0})$ is given by
\begin{align}\label{Equ:Dtaux0}
&D(\tau,x_{0})\notag\\
&=\sqrt{\left(\frac{2}{K(f)}W\left[\frac{K(f)}{2}\sqrt{\frac{\varrho\;\tau}
{P_{r}(x_{0})-\tau\sigma^{2}}}\right]\right)^2-{\left(h_{\textrm{A}}-h_{\textrm{U}}\right)}^2}.
\end{align}
Furthermore, we express $p_{c,\textrm{L}}(x_{0})$ as $p_{c,\textrm{L}}(x_{0})=\mathbb{P}\left[x_i\geq D(\tau, x_{0})\right]$, $\forall~i$, where AP$_i \in\Phi$. By defining  $\Phi_{c}$ as the set of APs contributing to the interference at the tagged UE which satisfy the condition $x_i\leq D(\tau, x_{0})$, we obtain $p_{c,\textrm{L}}(x_{0})$ as
\begin{align}\label{Equ:pc}
p_{c,\textrm{L}}(x_{0})=\mathbb{P}\left[n(\Phi_{c})=0\right].
\end{align}

We next calculate $\mathbb{P}\left[n(\Phi_{c})=0\right]$. We note that the location of the interferers follows a homogeneous PPP with the density $\lambda_{\textrm{A}}$. To determine the APs which belong to $\Phi_{c}$, its density needs to be found out. Thus, by considering \eqref{Equ:CondX}, \eqref{Equ:pep}, and \eqref{Equ:pc}, we evaluate the process where the APs belong to $\Phi_{c}$ as a probabilistic thinning of the original one, with the average density given by
\begin{align}\label{Equ:density1}
\Lambda_{\Phi_{c}}(x_{0})&=\int_{r_{\textrm{min}}}^{\hat{x}_{0}}
\int_{\theta_{0}-\frac{\varphi_{\textrm{U,H}}}{2}}^{\theta_{0}+\frac{\varphi_{\textrm{U,H}}}{2}} \lambda_{\textrm{A}}p_{\textrm{H,H}}(x)p_{\textrm{H,V}}(x)p_{\textrm{L}}(x) x d\theta dx\notag \\
&=\int_{r_{\textrm{min}}}^{\hat{x}_{0}}\int_{0}^{\varphi_{\textrm{U,H}}}\frac{\lambda_{\textrm{A}} \varphi_{\textrm{A,H}}}{2\pi}p_{\textrm{H,V}}(x)\zeta e^{-\eta x} x d\theta dx\notag\\
&=\frac{\lambda_{\textrm{A}} \zeta \varphi_{\textrm{A,H}} \varphi_{\textrm{U,H}}}{2\pi} \int_{r_{\textrm{min}}}^{\hat{x}_{0}} p_{\textrm{H,V}}(x) e^{-\eta x} x dx,
\end{align}
the integral in which can be calculated numerically. Here, we define $\hat{x}_{0}$ as $\hat{x}_{0}=\min\left\{D(\tau,x_{0}),r_{\textrm{max}}\right\}$. Thereafter, considering the void probability of the newly evaluated process, $p_{c,\textrm{L}}(x_{0})$ is derived as
\begin{align}\label{Equ:pcLF}
p_{c,\textrm{L}}(x_{0})=e^{-\Lambda_{\Phi_{c}}(x_{0})}.
\end{align}

Finally, by substituting \eqref{Equ:pL} and \eqref{Equ:pcLF} into \eqref{Equ:pc1}, the coverage probability for the link distance $x_{0}$ is derived as
\begin{align}\label{Equ:pcxF}
p_{c}(x_{0})&=p_{\textrm{L}}(x_{0})p_{c,\textrm{L}}(x_{0})\notag\\
&=\zeta e^{-\eta x_{0}} e^{-\Lambda_{\Phi_{c}}(x_{0})}
=e^{-\Omega(x_{0})},
\end{align}
where $\Omega(x_{0})=\Lambda_{\Phi_{c}}(x_{0})+\eta x_{0}+2\lambda_{\textrm{B}}r_{\textrm{B}}^2$.

\section{Numerical Results and Discussion}\label{sec:numerical}

In this section, we present numerical results for the coverage probabilities to examine the reliability performance of the considered THz communication system. The values of the parameters used in this section are summarized in Table \ref{tab1}, unless specified otherwise. Due to space limitation, in this section, we only present numerical results corresponding to a single narrowband that exist in the first transmission window above $1~\textrm{THz}$.  Also, we consider
$\varphi_{\Psi,\textrm{H}}=\varphi_{\Psi,\textrm{V}}$.

\begin{table}[t]
\caption{Value of System Parameters Used in Section~\ref{sec:numerical}}
\begin{center}
\begin{tabular}{|c|c|c|}
\hline
\textbf{Parameter} & \textbf{Symbol}& \textbf{Value} \\
\hline
Height of APs and UEs & $h_{\textrm{A}}$, $h_{\textrm{U}}$  & $3.0 $ m, $1.0 $ m\\
\hline
Height and radius of blockers & $h_{\textrm{B}}, r_{\textrm{B}}$ & $1.5 $ m, $0.3 $ m\\
\hline
Operating frequency and bandwidth  & $f$, $B$  & $1.07~\textrm{THz}$, $10~\textrm{GHz}$  \\ \hline
Absorption coefficient  \cite{2011_Jornet_TWC} & $K(f)$ &  $0.192$ $\textrm{m}^{-1}$  \\ \hline
Transmit power and AWGN power  & $P_{\textrm{T}}, \sigma^{2}$ &  $20~\textrm{dBm}, -74.4~\textrm{dBm}$ \\ \hline
Antenna gains of UEs and APs & $G_{\textrm{U}}$,$G_{\textrm{A}}$  &  $12.5~\textrm{dBi}$, $17.5~\textrm{dBi}$ \\ \hline
Densities of APs and blockers & $\lambda_{\textrm{A}}$, $\lambda_{\textrm{B}}$   &  0.1  m$^{-2}$, 0.2  m$^{-2}$  \\ \hline
\end{tabular}\label{tab1}
\end{center}\vspace{-3mm}
\end{table}

\begin{figure}[t]
\centering
\includegraphics[height=2.2in,width=0.95\columnwidth]{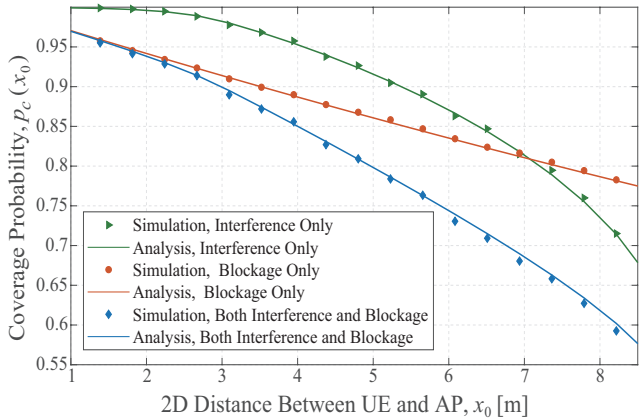}
\caption{Coverage probability versus the 2D UE-AP link distance for $\tau=3~\textrm{dB}$.}\label{Fig:Resplot1}
\end{figure}

Fig.~\ref{Fig:Resplot1} plots the coverage probabilities versus the 2D UE-AP link distance, $x_{0},$ for the SINR threshold of $\tau=3~\textrm{dB}$.
In this figure, we consider (i) coverage probability with both interference and blockage, (ii) coverage probability with interference only which is obtained by setting $p_{\textrm{L}}(x_{i})=1$, $\forall i$, and (iii) coverage probability with blockage only which is obtained by setting the density of interferers to zero. We first observe that the analytical results well match the simulation results, demonstrating the accuracy of our analytical results for the considered THz communication system. Second, we observe that the deterioration in coverage probability caused by interference is marginal for small $x_{0}$, but significantly increases when $x_{0}$ becomes large. This is due to the fact that when the UE is connected to a farther AP, in addition to the reduced received power, the impact of interference on the coverage probability becomes more detrimental since there are more interferers within the beamwidth of the UE. Third, we observe that the coverage probability with blockage only deteriorates when $x_{0}$ increases. This observation is expected since the effective number of blockers that exist in the UE-AP link increases with the distance of the link. These observations reveal that interference and blockage profoundly impact the coverage probability in THz communication systems; therefore, ignoring either of them leads to an overestimation of the system reliability, especially when $x_{0}$ is large.

\begin{figure}[t]
\centering
\includegraphics[height=2.2in,width=0.95\columnwidth]{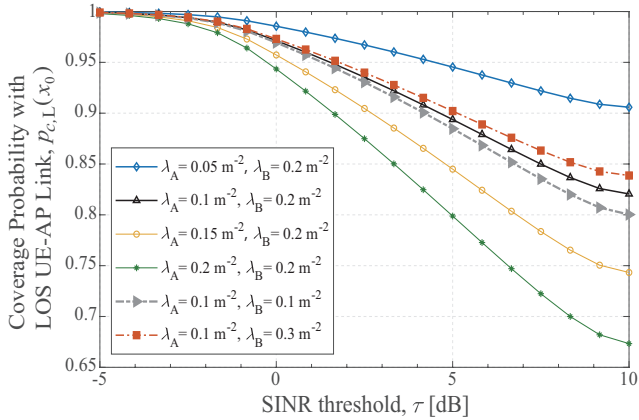}
\caption{Coverage probability with the LOS UE-AP link versus the SINR threshold for the UE-AP link distance of $5~\textrm{m}$.}\label{Fig:Resplot3}
\end{figure}

Fig.~\ref{Fig:Resplot3} plots the coverage probability when the UE-AP link of interest is in LOS, i.e., $p_{c,\textrm{L}}(x_{0})$ in \eqref{Equ:pcLF}, versus $\tau$, for different densities of APs and blockers when $x_{0}=5~\textrm{m}$. As expected, we first observe that $p_{c,\textrm{L}}(x_{0})$ becomes lower when $\tau$ increases.
Second, we observe that $p_{c,\textrm{L}}(x_{0})$ significantly decreases when the density of APs becomes higher, due to the increased impact from interferers. This demonstrates that network densification deteriorates the reliability of THz communication systems.
Third, we observe that $p_{c,\textrm{L}}(x_{0})$ improves when the density of blockers becomes higher. This is due to the fact that when there are more blockers, the likelihood of interference signals being blocked becomes higher, which leads
better $p_{c,\textrm{L}}(x_{0})$.

\begin{figure}[t]
\centering
\includegraphics[height=2.2in,width=0.95\columnwidth]{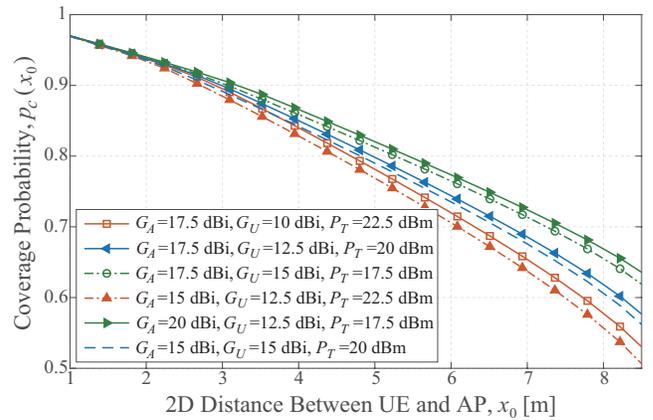}
\caption{Coverage probability versus the 2D UE-AP link distance for different antenna gains at APs and UEs.} \label{Fig:Resplot2}
\end{figure}

Fig.~\ref{Fig:Resplot2} plots the coverage probability versus $x_{0}$ for different antenna gains at APs and UEs, i.e., $G_{\textrm{A}}$ and $G_{\textrm{U}}$, for $\tau=3~\textrm{dB}$. Despite that different values of $G_{\textrm{A}}$ and $G_{\textrm{U}}$ are considered, in this figure we keep $P_{\textrm{T}}G_{\textrm{A}}G_{\textrm{U}}$ unchanged for the sake of fair comparison. First, we observe that the coverage probability becomes higher when $G_{\textrm{U}}$ increases. This is due to the fact that the beamwidths of the UEs become narrower when $G_{\textrm{U}}$ increases, which in turn decreases the number of interferers within the beamwidth of the UE, leading to less severe interference on the coverage performance. Second, we observe that the coverage probability improves when $G_{\textrm{A}}$ increases. The first and second observations reveal that the coverage performance of THz communication systems can be improved by increasing the antenna directivity at both the APs and the UEs. Finally, observing the curves with the same $P_{\textrm{T}}$, we find that the coverage probability improvement brought by increasing $G_{\textrm{A}}$ is higher than that brought by increasing $G_{\textrm{U}}$. This implies that it would be more worthwhile to increase the antenna directivity at the APs, rather than that at the UEs, to produce a more reliable THz communication system.

\section{Conclusions}\label{sec:conclusions}

We developed a novel easy-to-use analytical framework to investigate the reliability performance of 3D THz communication systems. Specifically, we derived new expressions for the coverage probability using dominant interferer analysis while considering the molecular absorption loss, 3D directional antennas at both UEs and APs, the interference from nearby APs, and the dynamic blockage caused by moving humans. Differing from the current THz studies, the proposed framework incorporates instantaneous interference and the vertical heights of THz devices. Using numerical results, we demonstrated the accuracy of our analysis and reveal useful insights into the impact of APs, blockers, AP-UE distance, and antenna directivity on the system coverage performance.

\appendices

\section{Derivation of $R_{\textrm{T}}$}\label{app:Derive_RT}

To find out the expression for $R_{\textrm{T}}$, we let the SNR when the UE-AP distance is $R_{\textrm{T}}$ equal the predefined threshold $\tau$. Therefore, we obtain
\begin{align}\label{Equ:DerRT}
\frac{\varrho \; e^{-K(f)\sqrt{(h_{\textrm{A}}-h_{\textrm{U}})^2+R_{\textrm{T}}^2}}}{((h_{\textrm{A}}-h_{\textrm{U}})^2+R_{\textrm{T}}^2) \sigma^{2}}=\tau.
\end{align}
By performing basic manipulation and using the definition of Lambert \textit{W}-function, we obtain
\begin{align}\label{Equ:DerRT_Result}
\frac{K(f)\sqrt{(h_{\textrm{A}}{-}h_{\textrm{U}})^2+R_{\textrm{T}}^2}}{2}= W\left[\frac{K(f)}{2}\sqrt{\frac{\varrho}{\tau\sigma^2}}\right].
\end{align}
By rearranging \eqref{Equ:DerRT_Result}, we arrive at \eqref{Equ:RT}.

\section{Derivation of $r_{\textrm{max}}$ and $r_{\textrm{min}}$}\label{app:Derive_rmax_rmin}

Let us focus on Fig. \ref{Fig:InfModel}. For $\psi_{0}\geq\frac{\varphi_{\textrm{U,V}}}{2}$, by observing the geometry of the spreading beam from the tagged UE, we obtain
\begin{align}\label{Equ:ramxproof1}
\tan\left(\frac{\pi}{2}-\psi_{0}\right)=\frac{x_{0}}{h_{\textrm{A}}-h_{\textrm{U}}}
\end{align} and
\begin{align}\label{Equ:ramxproof2}
\tan\left(\frac{\pi}{2}-\psi_{0}+\frac{\phi_{\textrm{U,V}}}{2}\right)
=\frac{r_{\textrm{max}}}{h_{\textrm{A}}-h_{\textrm{U}}}.
\end{align}
Also, for $\psi_{0}\leq\frac{\pi-\phi_{\textrm{U,V}}}{2}$, we obtain
\begin{align}\label{Equ:ramxproof3}
\tan\left(\frac{\pi}{2}-\psi_{0}-\frac{\phi_{\textrm{U,V}}}{2}\right)=\frac{r_{\textrm{min}}}{h_{\textrm{A}}-h_{\textrm{U}}}.
\end{align}
Then we expand \eqref{Equ:ramxproof2} and \eqref{Equ:ramxproof3} using the trigonometric properties given by $\tan(A\pm B) = (\tan(A) \pm \tan(B))/(1\mp \tan(A)\tan(B))$. Finally, by substituting \eqref{Equ:ramxproof1} into the expanded results, we arrive at \eqref{Equ:rmax} and \eqref{Equ:rmin}.

\section{Derivation of $p_{\textrm{H,V}}(x_{i})$}\label{app:Derive_15}

Let us denote $v_{i}$ as the distance of the link between AP$_{i}$ and its associating UE, and denote $\beta_{i}$ as the angle that the link between AP$_{i}$ and its associating UE form with the horizontal plane, as shown in the Fig. \ref{2a}.
For AP$_{i}$ to satisfy \textit{Condition 4)} stated in Section \ref{SubSec:phi}, $\beta_{i}$ needs to satisfy
\begin{equation}\label{Equ:beta_i}
\psi_{i}-\frac{\varphi_{\textrm{A,V}}}{2}\leq\beta_{i} \leq\psi_{i}+\frac{\varphi_{\textrm{A,V}}}{2}.
\end{equation}
Therefore, $p_{\textrm{H,V}}(x_{i})$ is obtained as
\begin{equation}\label{Equ:fbet}
p_{\textrm{\textrm{H,V}}}(x_{i})=\int_{\psi_{i}-\frac{\varphi_{\textrm{A,V}}}{2}}
^{\psi_{i}-\frac{\varphi_{\textrm{A,V}}}{2}}f_{\beta}(\beta_{i})d\beta_{i},
\end{equation}
where $f_{\beta}(\beta_{i})$ is the PDF of $\beta_{i}$.

To formulate $f_{\beta}(\beta_{i})$, we recall that UEs are distributed uniformly within the circle with radius $R_{\textrm{T}}$ centered at each AP and each AP associates with one UE only.
Therefore, the PDF of $v_{i}$, denoted by $f_{v}(v_{i})$, is expressed as
\begin{equation}\label{Equ:PDFv}
f_{v}(v_{i})=\begin{cases}
\frac{2 v_{i}}{R_{\textrm{T}}^2}, &0\leq v_{i}\leq R_{\textrm{T}},\\
0, &\textrm{otherwise}.
\end{cases}
\end{equation}
Then, by using the transformation $v_{i} =(h_{\textrm{A}}{-}h_{\textrm{U}})\cot(\beta_{i})$, we obtain
\begin{equation}\label{Equ:PDFbeta}
f_{\beta}(\beta_{i})=\begin{cases}
\frac{2(h_{\textrm{A}}{-}h_{\textrm{U}})^2}{R_{\textrm{T}}^2}\cot(\beta_{i})\csc^2(\beta_{i}), & \bar{\beta} \leq \beta_{i}\leq\frac{\pi}{2},\\
0, &\textrm{otherwise},
\end{cases}
\end{equation}
where $\bar{\beta} = \arctan\left(\frac{h_{\textrm{A}}{-}h_{\textrm{U}}}{R_{\textrm{T}}}\right)$. Finally, by substituting \eqref{Equ:PDFbeta} into \eqref{Equ:fbet} and solving the resultant integral by applying \cite[Eq (2.521)]{IntegralBook}, we obtain \eqref{Equ:pHV}.

\bibliographystyle{IEEEtran}
\bibliography{Akram_2020ICCWS_Ref}

\end{document}